\documentclass[conference]{IEEEtran}
\IEEEoverridecommandlockouts

\usepackage{amsmath,amssymb,amsfonts}
\usepackage{algorithmic}
\usepackage{graphicx}
\usepackage{textcomp}
\usepackage{xcolor}

\usepackage{tikz}
\usetikzlibrary{arrows.meta}
\allowdisplaybreaks
\usepackage[numbers, square, comma]{natbib}

\usepackage{hyperref}
\usepackage{cleveref}
\hypersetup{
    colorlinks=true, %set true if you want colored links
    linktoc=all,     %set to all if you want both sections and subsections linked
    linkcolor=black,  %choose some color if you want links to stand out
    citecolor=black
}

\newtheorem{theorem}{Theorem}
\newtheorem{lemma}{Lemma}
\newtheorem{definition}{Definition}
\newtheorem{remark}{Remark}

\def\BibTeX{{\rm B\kern-.05em{\sc i\kern-.025em b}\kern-.08em
    T\kern-.1667em\lower.7ex\hbox{E}\kern-.125emX}}

\begin{document}

\title{Covert Communication over Two Types of \\ Additive Noise Channels
}

\author{\IEEEauthorblockN{Cécile Bouette, Laura Luzzi, and Ligong Wang}
\IEEEauthorblockA{{ETIS, UMR 8051,} \\ {CY Cergy Paris Université, ENSEA, CNRS}\\
Cergy, France \\
Email: {\{cecile.bouette, laura.luzzi, ligong.wang\}@ensea.fr}}
}

\maketitle

\begin{abstract}
We extend previous results on covert communication over the additive white Gaussian noise channel to two other types of additive noise channels. The first is the Gaussian channel with memory, where the noise sequence is a Gaussian vector with an arbitrary invertible covariance matrix. We show that the fundamental limit for covert communication over such a channel is the same as over the channel with white, i.e., memoryless, Gaussian noise. The second type of channel we consider is one with memoryless generalized Gaussian noise. For such a channel we prove a general upper bound on the dominant term in the maximum number of nats that can be covertly communicated over $n$ channel uses. When the shape parameter $p$ of the generalized Gaussian noise distribution is in the interval $(0,1]$, we also prove a matching lower bound.
\end{abstract}

\section{Introduction}
We study the problem of “covert communication”, also known as “communication with low probability of detection” \cite{fundamental_covertness,bloch_resolvability,bash_article}, where the communicating parties do not want to let the eavesdropper detect whether transmission is taking place or not. Covertness is desirable in many of today's applications, for example, where even revealing \emph{who} is communicating, \emph{when}, and \emph{from where} can leak sensitive information.

For discrete memoryless channels (DMCs) and additive white Gaussian noise (AWGN) channels, the capacity (in nats per channel use) under a covertness constraint is equal to zero, because the maximum amount of information that can be transmitted reliably and covertly scales like the square-root of the total number of channel uses; this phenomenon is sometimes called the \textit{square-root law}.
The corresponding scaling constant $L$ was characterized in \cite{fundamental_covertness} for both DMCs and AWGN channels.

In this paper, we consider two separate extensions of the AWGN channel model. The first is to Gaussian channels with memory: the noise sequence is assumed to be a Gaussian stochastic process, and no longer independent and identically distributed (i.i.d.). 
We show that covert communication over such a channel is equivalent to that over an AWGN channel and, as a consequence, the square-root law holds, and the above-mentioned scaling constant $L$ remains the same irrespectively of the noise covariance matrix.

The second extension we consider is to memoryless additive noise channels whose noise has a generalized Gaussian distribution \cite{nadarajah2005,generalized_gaussian_distributions_short}. Such distributions are useful in modeling noise in many applications, e.g., noise spikes due to rare events \cite[Chapter 10]{fundamental_signal}, multiple-user interference in ultrawideband systems \cite{use_of_gg_distribution_interferences}, and atmospheric noise \cite{use_of_gg_distribution_atmospheric_noise}.
When the shape parameter $p$ of the generalized Gaussian distribution lies in $(0,1]$, we show that the square-root law continues to hold, and compute the exact value of the scaling constant $L$; for other values of $p$, we provide an upper bound on $L$.

The rest of this paper is arranged as follows: Section~\ref{sec:setup} describes the general setup of the problem; Section~\ref{sec:memory} studies Gaussian noise with memory; Section~\ref{sec:GG} studies generalized Gaussian noise; and Section~\ref{sec:future} briefly discusses possible directions for future works.

\section{General Setup}\label{sec:setup}
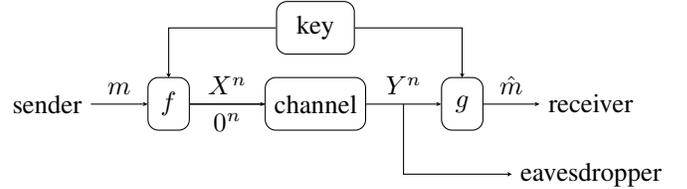
\begin{figure}[tbp]
\begin{center}
            \begin{tikzpicture}[
            nodetype1/.style={
                rectangle,
                rounded corners,
                minimum width=0.7cm,
                minimum height=0.7cm,
                draw=black,
                font=\normalsize
            },
            nodetype2/.style={
                rectangle,
                rounded corners,
                minimum width=0.55cm,
                minimum height=0.7cm,
                draw=black,
                font=\normalsize
            },
            tip2/.style={-{Stealth[length=0.6mm, width=0.5mm]}}
            ]
            \matrix[row sep=0.3cm, column sep=0.38cm, ampersand replacement=\&]{
            \& \& (invisible) \&\&
            \node (Key) [draw, nodetype1, text width=0.8cm, text centered]  {\text{key}}; \&\& \\
            \node (Alice) {\text{sender}};  \& \& \node (encoder) [nodetype2]   {$f$}; \&
            \node (X){}; \&
            \node (W) [draw, nodetype1, text width=1.1cm, text centered]  {\text{channel}}; \&
            \node (Y){};
            \&
            \node (decoder) [nodetype2] {$g$}; \&
            \node (Bob) {\text{receiver}};\\
            \& \& \&
            \& \& \& \&
            \node (Eve) {\text{eavesdropper}}; \& \\};
            
            \draw[->] (Alice) edge[tip2] node [above] {$m$} (encoder) ;
            \draw[->] (encoder) edge[tip2] node [above] (X) {} (W) ;
            \draw[-]  (encoder.east) -- node [above] (X1) {$X^n$}  node [below] (X2) {$0^n$} (W.west) ;
            \draw[arrows = {-Latex[length=1pt]}] (W) edge[tip2] node [above] {$Y^n$} (decoder) ;
            \draw[->] (decoder) edge[tip2] node [above] {$\hat{m}$} (Bob) ;
            \draw[-{Stealth[length=0.5mm, width=0.5mm]}] (Y.center)  |- node [above] {} (Eve) ;
            \draw[{Stealth[length=0.5mm, width=0.5mm]}-] (encoder) |-  (Key) ;
            \draw[-{Stealth[length=0.5mm, width=0.5mm]}] (Key) -| (decoder) ;
            \end{tikzpicture}
            \end{center}
    \caption{General setup for covert communications.}
     \label{fig:general_channel}
\end{figure}

The setup is illustrated in Fig.~\ref{fig:general_channel}. We consider an 
additive noise channel described by
\begin{equation}
\label{eq:channel}
    Y_i=X_i+Z_i,\qquad i=1,2,\ldots,
\end{equation}
where $X_i$ denotes the channel input, $Y_i$ the channel output, and $Z_i$ the additive noise, at time $i$, all of which take values in $\mathbb{R}$. 
We assume that the entire noise sequence is independent of the message and the secret key.

A deterministic code $\mathcal{C}$ of length $n$ for message set $\mathcal{M}$ consists of an encoder $f\colon\mathcal{M} \rightarrow \mathbb{R}^n , m \mapsto x^n$ and a decoder $g\colon \mathbb{R}^n \rightarrow \mathcal{M}, y^n \mapsto \hat{m}$.

The sender and the receiver are assumed to share a (sufficiently long) secret key, which is used to randomly select a code from the set of possible codes.
The eavesdropper is assumed to know the distribution used to select the code, but not the key or which code is effectively chosen for the transmission.

Covertness requires that the eavesdropper should not be able to detect whether a transmission is ongoing or not. Specifically, we consider the following covert communication requirement:
for some given $\delta >0$, the output distribution must satisfy
\begin{equation}
 \label{covert_communication_hypothesis}
     \mathbb{D}(P_{Y^n}||P_{Z^n}) \leq \delta,  %\quad\forall n
\end{equation}
where $\mathbb{D}(\cdot||\cdot)$ is the Kullback-Leibler divergence (relative entropy) \cite{coverthomas06}, $P_{Z^n}$ denotes the distribution of the noise vector $Z^n$, and $P_{Y^n}$ that of the output sequence \emph{averaged over the key} (i.e., over the randomly chosen code).

Given $\epsilon>0$, we denote by $K_n(\delta, \epsilon)$ the maximum of $\ln|\mathcal{M}|$ for which there exists a random code $\mathsf{C}$ of length $n$ that satisfies covertness condition \eqref{covert_communication_hypothesis}, and whose average probability of decoding error is at most $\epsilon$.
As in \cite{fundamental_covertness}, we define:
\begin{equation}
\label{L_definition}
L\triangleq \lim\limits_{\epsilon\downarrow  0} \varliminf\limits_{n\rightarrow  \infty} \dfrac{K_n(\delta,\epsilon)}{\sqrt{n\delta}}  .
\end{equation}

\section{Gaussian Noise With Memory}\label{sec:memory}
Consider the channel \eqref{eq:channel} 
where the noise sequence is a Gaussian process: for every $n$,
\begin{equation}
    \label{noise_with_memory}
    Z^n \sim \mathcal{N}(\boldsymbol{\mu}_n,\boldsymbol{\Sigma}_n),
\end{equation}
where the mean $\boldsymbol{\mu}_n$ is a length-$n$ vector, and where the covariance $\boldsymbol{\Sigma}_n$ is an $n\times n$ symmetric, positive definite matrix. Note that $\boldsymbol{\Sigma}_n$ being positive definite implies that it is invertible (i.e., non-singular).

We shall show that the fundamental limit for covert communication over this additive Gaussian noise channel with memory is the same as that over the AWGN channel: not only does $K_n$ grow like $\sqrt{n}$, but also $L=1$ as in \cite[Section V]{fundamental_covertness}. This result contrasts with the standard (non-covert) capacity of the Gaussian noise channel under an average-power constraint, which in general will change if AWGN is replaced by colored Gaussian noise \cite{coverthomas06}.

\vspace{2mm}

\begin{theorem}
\label{theorem_gaussian_channel_with_memory}
For the channel \eqref{eq:channel} with Gaussian noise \eqref{noise_with_memory}, under the covertness requirement \eqref{covert_communication_hypothesis},
\begin{equation}
L=1    
\end{equation}
irrespectively of $\boldsymbol{\mu}_n$ and $\boldsymbol{\Sigma}_n$.
\end{theorem}

\vspace{2mm}

\begin{IEEEproof}
We prove the theorem operationally by showing a one-to-one correspondence between codes for noise of the form \eqref{noise_with_memory} and i.i.d. noise. 

Since $\boldsymbol{\Sigma}_n$ is invertible, there exists an invertible $n \times n$ matrix $\mathbf{A}$ such that 
\begin{equation}
Z^n= \mathbf{A}\tilde{Z}^n+\boldsymbol{\mu}_n,
\end{equation}
where $\tilde{Z}^n$ is a standard Gaussian vector, i.e., it consists of i.i.d. entries $\mathcal{N}(0,1)$. Now consider the AWGN channel
\begin{equation}
\label{equivalent_channel}
\tilde{Y}_i = \tilde{X}_i + \tilde{Z}_i,\qquad i=1,\ldots,n.
\end{equation}
with time-$i$ input $\tilde{X}_i$ and output $\tilde{Y}_i$, respectively. Given any code $\mathcal{C}=(f,g)$ for the channel \eqref{eq:channel} with noise \eqref{noise_with_memory}, there is a corresponding code $\tilde{\mathcal{C}}=(\tilde{f},\tilde{g})$ for the AWGN channel \eqref{equivalent_channel}, and vice versa. Indeed, given $\mathcal{C}$, we construct $\tilde{\mathcal{C}}$ via:
    \begin{itemize}
        \item for all $m \in \mathcal{M}$,
        \begin{equation}\tilde{f}(m)= \mathbf{A}^{-1} f(m);
        \end{equation}
        \item for all $\tilde{y}^n \in \mathbb{R}^n$,
        \begin{equation}
        \tilde{g}\left(\tilde{y}^n\right)= g\left(\mathbf{A}\tilde{y}^n+\boldsymbol{\mu}_n\right).
        \end{equation}
    \end{itemize}
Reversely, given $\tilde{\mathcal{C}}$, we construct $\mathcal{C}$ via:
    \begin{itemize}
        \item for all $m \in \mathcal{M}$,
        \begin{equation}{f}(m)= \mathbf{A} \tilde{f}(m);
        \end{equation}
        \item for all $y^n \in \mathbb{R}^n$,
        \begin{equation}
        g\left(y^n\right)= \tilde{g}\left(\mathbf{A}^{-1}(y^n-\boldsymbol{\mu}_n)\right).
        \end{equation}
    \end{itemize}
By this construction, a decoding error occurs with code $\mathcal{C}$ on the channel \eqref{eq:channel} with \eqref{noise_with_memory} if, and only if, a decoding error occurs with code $\tilde{\mathcal{C}}$ on the channel \eqref{equivalent_channel}. Consequently, the error probabilities of the two codes (when used on their corresponding channels) are equal. This one-to-one correspondence applies to random codes on the two channels as well.
Furthermore, averaged over the random codes,
\begin{equation}\label{eq:Dequal}
\mathbb{D}(P_{{Y}^n}||P_{{Z}^n}) = \mathbb{D}(P_{\tilde{Y}^{n}}||P_{\tilde{Z}^{n}}),
\end{equation}
because the same invertible mapping---subtraction by $\boldsymbol{\mu}_n$ and then multiplication by $\mathbf{A}^{-1}$---maps $Y^n$ to $\tilde{Y}^n$ and $Z^n$ to $ \tilde{Z}^n$, hence the data-processing inequality for the Kullback-Leibler divergence \cite[Theorem 2.15]{info_theory_polyanskiy} holds in both directions.

We have now shown that the corresponding random codes on the two channels have exactly the same error probability and covertness property. The theorem then follows because, by \cite[Theorem 5]{fundamental_covertness}, $L=1$ for the channel \eqref{equivalent_channel}.
\end{IEEEproof}

\section{Memoryless Generalized Gaussian Noise}\label{sec:GG}
We again consider the additive noise channel described by \eqref{eq:channel}, but we now assume that $Z^n$ is i.i.d., with every entry having a generalized Gaussian distribution \cite{nadarajah2005,generalized_gaussian_distributions}: for some $p>0, \alpha>0$, the probability density function $f_Z$ of $Z$ is
\begin{equation}
\label{noise}
f_Z(z)=\frac{c_p}{\alpha}e^{-\frac{|z|^p}{2\alpha^p}},\qquad z\in\mathbb{R},
\end{equation}
where
\begin{equation}
c_p=\frac{p}{2^{\frac{p+1}{p}}\Gamma(\frac{1}{p})},
\end{equation}
with $\Gamma(\cdot)$ denoting the gamma function. We denote $Z\sim\mathcal{N}_p(0,\alpha^p)$ for simplicity. Note that:
\begin{IEEEeqnarray}{rCl}
    \mathbb{E}[|Z|^p]&=&\frac{2\alpha^p}{p},\\
    \label{entropy_gg}
    h(Z)&=&\ln\left(\frac{\alpha}{c_p}\right) +\frac{1}{p},
\end{IEEEeqnarray}
where $h(\cdot)$ denotes the differential entropy \cite{coverthomas06}.

\subsection{An upper bound}
\begin{theorem}
\label{theorem_converse}
For the channel \eqref{eq:channel} with memoryless generalized Gaussian noise \eqref{noise}, under the covertness condition \eqref{covert_communication_hypothesis},
\begin{equation}
L\leq\sqrt{\frac{2}{p}}
\end{equation}
irrespectively of the parameter $\alpha$.
\end{theorem}
\vspace{2mm}

Before proving Theorem \ref{theorem_converse}, we present a lemma.

\begin{lemma}
\label{inequalities_entropy_dk_gg}
For any real random variable $Y$, let $\gamma\in\mathbb{R}$ be such that $ \gamma^p=\frac{p}{2}\mathbb{E}[|Y|^p]$. Then we have the two following inequalities:
\begin{IEEEeqnarray}{rCl}
\label{entropy_upper_bound}
    h(Y)&\leq& \ln\left(\frac{\gamma}{c_p}\right)+\frac{1}{p}\\
\label{relative_divergence_upper_bound}
    \mathbb{D}(P_{Y}||P_{Z})&\geq&\ln\left(\frac{\alpha}{\gamma}\right)+\frac{1}{p}\left(\frac{\gamma^p}{\alpha^p}-1\right).
\end{IEEEeqnarray}
\end{lemma}
Equality holds in both \eqref{entropy_upper_bound} and \eqref{relative_divergence_upper_bound}  if $Y\sim\mathcal{N}_p(0,\gamma^p)$.

\begin{IEEEproof}
Let $\tilde{Z}\sim\mathcal{N}_p(0,\gamma^p)$.
First we show \eqref{entropy_upper_bound} via the following:
\begin{IEEEeqnarray}{rCl}
\label{maximisation_entropy}
    0&\leq&\mathbb{D}(P_{Y}||P_{\tilde{Z}})\nonumber\\
    &=&-h(Y)-\int_{\mathbb{R}}f_{Y}(y)\ln\left(f_{\tilde{Z}}(y)\right)\textnormal{d}y\nonumber\\
    &=&-h(Y)-\int_{\mathbb{R}}f_{Y}(y)\ln\left(\frac{c_p}{\gamma} e^{-\frac{|y|^p}{2\gamma^p}}\right)\textnormal{d}y\nonumber\\
    &=&-h(Y)-\ln\left(\frac{c_p}{\gamma}\right)+\frac{\mathbb{E}[|Y|^p]}{2\gamma^p}\nonumber\\
    &=&-h(Y)-\ln\left(\frac{c_p}{\gamma}\right)+\frac{1}{p}.
\end{IEEEeqnarray}
From \eqref{maximisation_entropy}, we immediately obtain \eqref{entropy_upper_bound}, and that equality is achieved when $Y$ has the same distribution as $\tilde{Z}$.

We then show \eqref{relative_divergence_upper_bound} via the following:
\begin{IEEEeqnarray}{rCl}
    \mathbb{D}(P_{Y}||P_Z)&=&-h(Y)-\int_{\mathbb{R}}f_{Y}(y)\ln(f_Z(y))\textnormal{d}y\nonumber\\
    &=&-h(Y)-\int_{\mathbb{R}}f_{Y}(y)\ln\left(\frac{c_p}{\alpha} e^{-\frac{|y|^p}{2\alpha^p}}\right)\textnormal{d}y\nonumber\\
    &=&-h(Y)-\ln\left(\frac{c_p}{\alpha}\right)+\frac{\mathbb{E}[|Y|^p]}{2\alpha^p}\nonumber\\
    \label{use_of_first_inequality}
    &\geq&-\ln\left(\frac{\gamma}{c_p}\right)-\frac{1}{p}-\ln\left(\frac{c_p}{\alpha}\right)+\frac{\mathbb{E}[|Y|^p]}{2\alpha^p}\IEEEeqnarraynumspace\\
    &=&\ln\left(\frac{\alpha}{\gamma}\right)-\frac{1}{p}+\frac{\gamma^p}{p\alpha^p}
\end{IEEEeqnarray}
which is the desired inequality. Note that \eqref{use_of_first_inequality} follows from \eqref{entropy_upper_bound}, and that it holds with equality when $Y$ has the same distribution as $\tilde{Z}$.
\end{IEEEproof}
\vspace{2mm}
\begin{IEEEproof}[Proof of Theorem \ref{theorem_converse}]
Take any random code $\mathsf{C}$ of length $n$.
Let $\bar{P}_{X}$ and $\bar{P}_{Y}$ denote the  average input and output distributions over all possible codes, an uniformly drawn message, and the $n$ channel uses.
Notice that $\bar{P}_{Y}$ is the output distribution corresponding to input distribution $\bar{P}_{X}$.

Starting with the condition \eqref{covert_communication_hypothesis}, similarly to \cite{fundamental_covertness} we have:
\begin{IEEEeqnarray}{rCl}
\label{block_chain_divergence}
    \delta&\geq&\mathbb{D}(P_{Y^n}||P_{Z^n})\nonumber\\
    &=&-h(Y^n)-\mathbb{E}[\ln \left(f_{Z^n}(Y^n)\right)]\nonumber\\
    &=&\sum_{i=1}^n\left(-h(Y_i|Y^{i-1})-\mathbb{E}[\ln(f_{Z}(Y_i))]\right)\nonumber\\
    &\geq&\sum_{i=1}^n\left(-h(Y_i)-\mathbb{E}[\ln(f_{Z}(Y_i))]\right)\nonumber\\
    &=&\sum_{i=1}^n\mathbb{D}(P_{Y_i}||P_{Z})\nonumber\\
    &\geq& n\,\mathbb{D}(\bar{P}_{Y}||P_{Z}),
\end{IEEEeqnarray}
where the last step follows because the Kullback-Leibler divergence is convex.
Let
\begin{equation}
    \gamma_n\triangleq\left(\frac{p}{2}\mathbb{E}_{\bar{P}_Y}[|Y|^p]\right)^{\frac{1}{p}},
\end{equation}
then by \eqref{relative_divergence_upper_bound} in Lemma \ref{inequalities_entropy_dk_gg} and \eqref{block_chain_divergence}:
\begin{IEEEeqnarray}{rCl}
    \label{gamma_n_upper_bound_preliminary}
    \ln\left(\frac{\alpha}{\gamma_n}\right)+\frac{1}{p}\left(\frac{\gamma_n^p}{\alpha^p}-1\right)\leq\frac{\delta}{n}.
\end{IEEEeqnarray}
As $n\rightarrow \infty$, the left-hand side of \eqref{gamma_n_upper_bound_preliminary} must approach zero, which requires that
\begin{equation}
    \lim_{n\to\infty} \gamma_n = \alpha.
\end{equation}
Notice that:
\begin{IEEEeqnarray}{rCl}
\label{limit_divergence}
    \lim\limits_{\gamma_n\rightarrow \alpha}\dfrac{\ln\left(\frac{\alpha}{\gamma_n}\right)+\frac{1}{p}\left(\frac{\gamma_n^p}{\alpha^p}-1\right)}{\left(\frac{\gamma_n}{\alpha}-1\right)^2}=\frac{p}{2}.
\end{IEEEeqnarray}
Therefore \eqref{gamma_n_upper_bound_preliminary} implies:
\begin{IEEEeqnarray}{rCl}
    \label{gamma_n_upper_bound}
   \varlimsup\limits_{n\rightarrow \infty}\dfrac{\frac{\gamma_n}{\alpha}-1}{\sqrt{\frac{\delta}{n}}}\leq\sqrt{\frac{2}{p}}.
\end{IEEEeqnarray}
In other words,
\begin{IEEEeqnarray}{rCl}
        \left(\frac{\gamma_n}{\alpha}-1\right)\leq\sqrt{\frac{2}{p}\frac{\delta}{n}}+o\left(\frac{1}{\sqrt{n}}\right).
\end{IEEEeqnarray}

For each realization $\mathcal{C}$ of the random code $\mathsf{C}$, we denote by $\epsilon_n(\mathcal{C})$ its error probability. Let $\epsilon_n$ be the average error probability over the random codebook. For each $\mathcal{C}$, we have by Fano's inequality:
\begin{equation}   \ln\left|\mathcal{M}\right|(1-\epsilon_n(\mathcal{C}))- 1  \leq I(X^n;Y^n|\mathsf{C}=\mathcal{C}).
\end{equation}
By averaging over the random code, we obtain
\begin{IEEEeqnarray}{rCl}
   \ln\left|\mathcal{M}\right|(1-\epsilon_n)- 1
    &\leq&I(X^n;Y^n|\mathsf{C})\nonumber\\
    & \leq &I(X^n,\mathsf{C}; Y^n)\nonumber\\
    \label{markov_chain_inequality}
    &=&I(X^n;Y^n)\\
   &=& \sum_{i=1}^n I(X^n ; Y_i |Y^{i-1} )\nonumber\\
  &=& \sum_{i=1}^n \left(h(Y_i |Y^{i-1} ) -h(Y_i |X^n, Y^{i-1} )\right)\nonumber \\
  &=& \sum_{i=1}^n \left(h(Y_i|Y^{i-1}) - h(Y_i |X_i)\right)\nonumber\\
  &\leq& \sum_{i=1}^n I(X_i;Y_i)\nonumber\\
  \label{mutual_information_upper_bound_first_tmp}
  &\leq& nI( \bar{P}_{X},W),
\end{IEEEeqnarray}
where $W$ denotes the channel law corresponding to \eqref{eq:channel} with \eqref{noise}; \eqref{markov_chain_inequality} holds by the Markov chain:
\begin{equation}
\label{markov_chain}
    \mathsf{C}\rightarrow X^n \rightarrow Y^n;
\end{equation}
and \eqref{mutual_information_upper_bound_first_tmp} holds because mutual information is concave in the input distribution.
By the definition of $K_n(\delta,\epsilon)$, \eqref{mutual_information_upper_bound_first_tmp} implies 
\begin{IEEEeqnarray}{rCl}
\label{mutual_information_upper_bound_first}
    K_n(\delta,\epsilon_n)(1-\epsilon_n)- 1 &\leq& nI( \bar{P}_{X},W).
\end{IEEEeqnarray}
By \eqref{entropy_upper_bound} in Lemma \ref{inequalities_entropy_dk_gg} and \eqref{entropy_gg}, we know that:
\begin{IEEEeqnarray}{rCl}
    I( \bar{P}_{X},W)&=&h(\bar{P}_{Y})-h(Z)\nonumber\\
    &\leq& \ln\left(\frac{\gamma_n}{c_p}\right)+\frac{1}{p} -h(Z)\nonumber\\
    &=& \ln\left(\frac{\gamma_n}{\alpha}\right)\nonumber\\
    \label{mutual_information_upper_bound}
    &\leq& \frac{\gamma_n}{\alpha}-1.
\end{IEEEeqnarray}
Combining \eqref{mutual_information_upper_bound_first} and \eqref{mutual_information_upper_bound} we have:
\begin{equation}
\label{mutual_information_upper_bound_second}
    K_n(\delta,\epsilon_n)(1-\epsilon_n)- 1\leq n\left(\frac{\gamma_n}{\alpha}-1\right).
\end{equation}

Finally, by combining \eqref{mutual_information_upper_bound_second} and \eqref{gamma_n_upper_bound} and recalling the definition of $L$ in \eqref{L_definition} we obtain the desired result.
\end{IEEEproof}

\subsection{A matching lower bound for $p\in(0,1]$}

It is evident from \cite{fundamental_covertness} that the upper bound of Theorem \ref{theorem_converse} is tight for $p=2$. We show that it is also tight when $0< p\leq 1$.

\begin{theorem}
\label{theorem_achievability}
For the additive noise channel \eqref{eq:channel} with memoryless generalized Gaussian noise \eqref{noise}, under the covertness condition \eqref{covert_communication_hypothesis},
\begin{equation}
L=\sqrt{\frac{2}{p}},\qquad 0<p\le 1,
\end{equation}
irrespectively of the noise parameter $\alpha$.
\end{theorem}

\vspace{2mm}

Before proving Theorem \ref{theorem_achievability}, we recall the definition of self decomposability and a relevant lemma.

\begin{definition}[{\cite{generalized_gaussian_distributions,lukacs}}]
\label{definition_self_decomposability}
    A generalized Gaussian distribution $\mathcal{N}_p(0, \alpha^p)$ is self decomposable if, for every
$\beta \geq 1$, there exists a random variable $V_\beta$ such that
\begin{equation}
    \beta Z = V_\beta + U
\end{equation}
where $Z, U \sim\mathcal{N}_p(0, \alpha^p)$ and $U$ is independent of $V_\beta$.
\end{definition}

\begin{lemma}[{\cite[Theorem 6]{generalized_gaussian_distributions}}]
\label{lemma_self_decomposability}
    The distribution $\mathcal{N}_p(0,\alpha^p)$ is self decomposable for every $p \in (0,1]$.
\end{lemma}

\vspace{2mm}
\begin{IEEEproof}[Proof of Theorem \ref{theorem_achievability}]
The converse part of Theorem \ref{theorem_achievability} comes directly from Theorem \ref{theorem_converse}. The achievability part follows a similar argument as \cite[Section V-B]{fundamental_covertness}, as we detail below.
For a total of $n$ channel uses, let
\begin{equation}
\label{choice_gamma_n}
\gamma_n=\alpha\left(1+\sqrt{2p\frac{\delta}{n}}\right)^{\frac{1}{p}}.
\end{equation}
We consider a random code in which every codeword is independent of every other codeword and i.i.d. according to $P_X$, which satisfies
\begin{equation}
    \label{definition_self_decomposable}
    (X+Z)\sim \mathcal{N}_p(0,\gamma_n^p),
\end{equation}
where $Z\sim\mathcal{N}_p(0,\alpha^p)$ is independent of $X$. The existence of such a $P_X$ is guaranteed by the self-decomposability property from Lemma~\ref{lemma_self_decomposability}.
We denote by $X^n=(X_1,\dots,X_n)$ the associated i.i.d. input sequence, and by $Y^n=(Y_1,\dots,Y_n)$ the corresponding  output sequence, with $Y_i\sim\mathcal{N}_p(0,\gamma_n^p)$ for all $1\leq i\leq n$. Note that the distribution of every input and output symbol depends on $n$.

We check that this random code satisfies the covertness condition  \eqref{covert_communication_hypothesis}:
\begin{IEEEeqnarray}{rCl}
\mathbb{D}(P_{Y^n}||P_{Z^n})
    &=&n\hspace{01mm}\mathbb{D}(P_{Y}||P_{Z})\label{eq:DPYZn}\\
    \label{divergence_expression}
    &=& n\left(\ln\left(\frac{\alpha}{\gamma_n}\right)+\frac{1}{p}\left(\frac{\gamma_n^p}{\alpha^p}-1\right)\right)\IEEEeqnarraynumspace \\
    \label{lower_bound_ln}
    &\leq& n\frac{1}{2p}\left(\frac{\gamma_n^p}{\alpha^p}-1\right)^2\\
    \label{covert_communication_hypothesis_verification}
    &=& \delta,
\end{IEEEeqnarray}
where \eqref{eq:DPYZn} follows because both $Y^n$ and $Z^n$ are i.i.d.; \eqref{divergence_expression} because we have equality in \eqref{relative_divergence_upper_bound}; and \eqref{lower_bound_ln} because $\ln(1+a)\geq a -\frac{a^2}{2}$, $a>-1$.

We shall show that, for the above joint distribution,
\begin{equation}
\label{lower_boud_L}
    \lim\limits_{\epsilon\downarrow  0}\varliminf\limits_{n \rightarrow \infty} \frac{K_n(\delta,\epsilon)}{\sqrt{n}} \geq \lim\limits_{n \rightarrow \infty} \frac{1}{\sqrt{n}} I(X^n;Y^n).
\end{equation}
From \cite[Section II]{general_capacity} we know that for all $n\in\mathbb{N}^+$, $\epsilon>0$, and $\gamma>0$, there exists a code of length $n$ with message alphabet $\mathcal{M}$, whose error probability  $\epsilon$ is bounded as
\begin{equation}
\label{feinstein_lemma}
    \epsilon\leq\mathbb{P}\left\{\frac{i_{X^n,Y^n}(X^n,Y^n)}{n} \leq \frac{\ln|\mathcal{M}|}{n} +\gamma\right\}+e^{-\gamma n},
\end{equation}
where
\begin{equation}
\label{definition_information_density}
    i_{X^n,Y^n}(x^n,y^n)=\ln\left(\frac{f_{Y^n|X^n}(y^n|x^n)}{f_{Y^n}(y^n)}\right)
\end{equation}
is the information density.
Choosing $\gamma=n^{-\frac{3}{4}}$ and $\epsilon= 2e^{-n^{\frac{1}{4}}}$, \eqref{feinstein_lemma} becomes:
\begin{equation}
\label{feinstein_lemma_step}
    \frac{\epsilon}{2}\leq\mathbb{P}\left\{\frac{i_{X^n,Y^n}(X^n,Y^n)}{\sqrt{n}} \leq \frac{\ln|\mathcal{M}|}{\sqrt{n}} +n^{-\frac{1}{4}}\right\}.
\end{equation}
Letting $n\to\infty$ in \eqref{feinstein_lemma_step}, we have that there exists a sequence of codes with error probabilities $\epsilon_n=2e^{-n^{\frac{1}{4}}}$ such that:
\begin{IEEEeqnarray}{rCl}
\label{fundamental_formula_capacity}
\varliminf\limits_{n \rightarrow \infty} \frac{\ln|\mathcal{M}|}{\sqrt{n}}
     &\geq& \mathbb{P}\text{-}\liminf\limits_{n \rightarrow \infty} \frac{1}{\sqrt{n}} i_{X^n,Y^n}(X^n,Y^n),
\end{IEEEeqnarray}
where $\mathbb{P}\text{-}\liminf$ denotes the \textit{limit inferior in probability}; see \cite{general_capacity}.
We next study the right-hand side of \eqref{fundamental_formula_capacity}. We shall show that the term inside the $\mathbb{P}\text{-}\liminf$ converges in probability towards its expectation 
\begin{equation}
\label{expection_information_density}
\mathbb{E}\left[\frac{1}{\sqrt{n}} i_{X^n,Y^n}(X^n,Y^n)\right] = \frac{1}{\sqrt{n}} I(X^n;Y^n).
\end{equation}
By Chebyshev's inequality, for any $a >0$,
\begin{IEEEeqnarray}{rCl}
\IEEEeqnarraymulticol{3}{l}{
\mathbb{P}\left\{\left|\frac{1}{\sqrt{n}} i_{X^n,Y^n}(X^n,Y^n)- \mathbb{E}\left[ \frac{1}{\sqrt{n}} i_{X^n,Y^n}(X^n,Y^n)\right]\right| \geq a\right\}
}\nonumber\\* \quad
\label{chebyshev}
~~~~~~~~~~~~~~~~~~~~&\leq& \frac{\mathrm{var}\left(\dfrac{1}{\sqrt{n}} i_{X^n,Y^n}(X^n,Y^n)\right)}{a^2}.
\end{IEEEeqnarray}
Hence, to prove the desired convergence, it suffices to show that the variance on the right-hand side of \eqref{chebyshev} converges to $0$, which is verified as follows:
\begin{IEEEeqnarray}{rCl}
    \IEEEeqnarraymulticol{3}{l}{
    \mathrm{var}\left(\frac{1}{\sqrt{n}} i_{X^n,Y^n}(X^n,Y^n)\right)
    }\nonumber\\*
    &=&\mathrm{var}\left(\frac{1}{\sqrt{n}} \ln\left(\frac{f_{Y^n|X^n}(Y^n|X^n)}{f_{Y^n}(Y^n)}\right)\right)\nonumber\\
    &=&\mathrm{var}\left(\frac{1}{\sqrt{n}}\ln\left(\frac{\prod_{i=1}^n \frac{c_p}{\alpha} e^{-\frac{|Z_i|^p}{2\alpha^p}}}{\prod_{i=1}^n \frac{c_p}{\gamma_n} e^{-\frac{|Y_i|^p}{2\gamma_n^p}}}\right)\right)\nonumber\\
    &=&\frac{1}{n}\mathrm{var}\left( -\sum_{i=1}^n\frac{|Z_i|^p}{2\alpha^p}+\sum_{i=1}^n\frac{|Y_i|^p}{2\gamma_n^p}\right)\nonumber\\
    &=&\frac{1}{n}\sum_{i=1}^n \mathrm{var}\left( \frac{|Y_i|^p}{2\gamma_n^p}-\frac{|Z_i|^p}{2\alpha^p}\right)\nonumber\\
    &=&\mathrm{var}\left( \frac{|Y|^p}{2\gamma_n^p}-\frac{|Z|^p}{2\alpha^p}\right)\nonumber\\
    &=&\frac{1}{4\gamma_n^{2p}\alpha^{2p}}\mathbb{E}\bigl[(|\alpha Y|^p-|\gamma_nZ|^{p})^2\bigr]\nonumber\\
    &&{}-\frac{1}{4\gamma_n^{2p}\alpha^{2p}}\bigl(\mathbb{E}\left[|\alpha Y|^p-|\gamma_nZ|^{p}\right]\bigr)^2\nonumber\\
    &=&\frac{1}{4\gamma_n^{2p}\alpha^{2p}}\mathbb{E}\bigl[(|\alpha Y|^p-|\gamma_nZ|^{p})^2\bigr]\nonumber\\
    \label{concavity}
    &\leq&\frac{1}{4\gamma_n^{2p}\alpha^{2p}}\mathbb{E}\bigl[(|\alpha Y-\gamma_nZ|^p)^2\bigr]\\
    &=&\frac{1}{4\gamma_n^{2p}\alpha^{2p}}\mathbb{E}\bigl[|( \alpha X+(\alpha-\gamma_n)Z)^2|^p\bigr]\nonumber\\
    &=&\frac{1}{4\gamma_n^{2p}\alpha^{2p}}\mathbb{E}\bigl[|\alpha^2 X^2+(\alpha-\gamma_n)^2 Z^2 +2\alpha(\alpha-\gamma_n) XZ|^p\bigr]\nonumber\\
   % \label{jensen_concavity}
    &\leq&\frac{1}{4\gamma_n^{2p}\alpha^{2p}}\Big|\mathbb{E}[\alpha^2 X^2+(\alpha-\gamma_n)^2 Z^2 +2\alpha (\alpha-\gamma_n) XZ]\Big|^p\nonumber\\*
    \ \label{jensen_concavity}\\
    &=&\frac{1}{4\gamma_n^{2p}\alpha^{2p}}\Big|\mathbb{E}\left[\alpha^2 X^2\right]+(\alpha-\gamma_n)^2 \mathbb{E}\left[Z^2\right]\Big|^p\nonumber\\
    &=&\frac{1}{4\gamma_n^{2p}\alpha^{2p}}\Bigg|\alpha \frac{2^{\frac{2}{p}}(\gamma_n^2-\alpha^2)}{\Gamma(\frac{1}{p})}\Gamma\left(\frac{3}{p}\right)\nonumber\\*
    && \qquad\qquad\quad{}+(\alpha-\gamma_n)^2 \frac{2^{\frac{2}{p}}\alpha^2}{\Gamma(\frac{1}{p})}\Gamma\left(\frac{3}{p}\right)\Bigg|^p,
    \label{variance_p_under_1}
\end{IEEEeqnarray}
where \eqref{concavity} holds because, for all $a,b\in\mathbb{R}$ and $0<p\leq 1$, $||a|^p-|b|^p|\leq|a-b|^p$ due to the concavity of $t\mapsto|t|^p$; and \eqref{jensen_concavity} holds by Jensen's inequality. Since, by our choice, $\gamma_n\rightarrow\alpha$ when $n\rightarrow\infty$, the right-hand side of \eqref{variance_p_under_1} approaches zero. This establishes \eqref{expection_information_density} and, moreover,
\begin{equation}
    \mathbb{P}\text{-}\liminf\limits_{n \rightarrow \infty} \frac{1}{\sqrt{n}} i_{X^n,Y^n}(X^n,Y^n) = \varliminf\limits_{n\rightarrow \infty}\frac{I(X^n;Y^n)}{\sqrt{n}},
\end{equation}
which concludes the proof of \eqref{lower_boud_L}.
We continue from \eqref{lower_boud_L} to complete the proof:
\begin{IEEEeqnarray}{rCl}
    \IEEEeqnarraymulticol{3}{l}{
    \lim\limits_{\epsilon\downarrow  0} \varliminf\limits_{n \rightarrow \infty} \frac{K_n(\delta, \epsilon)}{\sqrt{n}\sqrt{\delta}}
    }\nonumber\\* \quad
    &\geq&\varliminf\limits_{n\rightarrow \infty}\dfrac{I(X^n,Y^n)}{\sqrt{n}\sqrt{\delta}}\nonumber\\
    &=&\lim\limits_{n\rightarrow \infty}\frac{\sqrt{n}}{\sqrt{\delta}}I(X,Y)\nonumber\\
    &=&\varliminf\limits_{n\rightarrow \infty}\frac{\sqrt{n}}{\sqrt{\delta}}\left(h(Y)-h(Z)\right)\nonumber\\
    &=&\lim\limits_{n\rightarrow \infty}\frac{\sqrt{n}}{\sqrt{\delta}}\left(\ln\left(\frac{\gamma_n}{c_p}\right)+\frac{1}{p}
    -\ln\left(\frac{\alpha}{c_p}\right)-\frac{1}{p}\right)\nonumber\\
    &=&\lim\limits_{n\rightarrow \infty}\frac{\sqrt{n}}{\sqrt{\delta}}\ln\left(\frac{\gamma_n}{\alpha}\right)\nonumber\\
    &=&\sqrt{\frac{2}{p}},
\end{IEEEeqnarray}
which is the desired lower bound.
\end{IEEEproof}
\vspace{2mm}
\begin{remark}
Theorems~\ref{theorem_converse} and~\ref{theorem_achievability} continue to hold when the noise sequence is of the form
\begin{equation}\label{eq:GGmemory}
Z^n = \mathbf{A} \tilde{Z}^n,
\end{equation}
where $\mathbf{A}$ is an invertible $n\times n$ real matrix, and $\tilde{Z}^n$ is i.i.d. with each entry having a generalized Gaussian distribution. This can be shown by applying the same proof techniques that we used in proving Theorem~\ref{theorem_gaussian_channel_with_memory}. Such noise distributions are called ``multivariate $\theta$-generalized normal distributions'' in some literature \cite{mutivariate_gg_def}.
\end{remark}

\section{Future Work}\label{sec:future}
For memoryless generalized Gaussian noise with shape parameter $p>1$, $p\neq 2$, it remains to find a lower bound to match the upper bound of Theorem~\ref{theorem_converse} (if that upper bound is tight). Additionally, one could explore whether or not our proof techniques can be applied to more general noise distributions.

\section*{Acknowledgment}

This work was supported in part by CY Initiative of Excellence (grant \textit{Investissements d'Avenir} ANR-16-IDEX-0008).

\newpage
\bibliographystyle{IEEEtran}
\begin{small}
\bibliography{biblio}
\end{small}

\end{document}